\documentclass[nanomaterials,article,accept,moreauthors,pdftex]{Definitions/mdpi_arxiv} 

\firstpage{1} 
\pubvolume{1}
\issuenum{1}
\articlenumber{0}
\pubyear{2021}
\copyrightyear{2020}
\datereceived{} 
\dateaccepted{} 
\datepublished{} 
\hreflink{https://doi.org/} 

\usepackage{subcaption}				
\usepackage[noabbrev]{cleveref}
\usepackage[retain-unity-mantissa=false]{siunitx} 
\usepackage{physics}
\usepackage{color}

\newcommand{\el}{_\text{el}}
\newcommand{\ph}{_\text{ph}}
\newcommand{\elel}{_\text{el*-el}}
\newcommand{\elph}{_\text{el*-ph}}
\newcommand{\pt}{_\text{pt}}
\Title{Implementation of the electronic non-equilibrium in the two-temperature model}

\TitleCitation{Implementation of the electronic non-equilibrium in the two-temperature model}

\Author{Markus Uehlein$^*$, Sebastian T. Weber and Baerbel Rethfeld}

\AuthorNames{Markus Uehlein, Sebastian T. Weber and Baerbel Rethfeld}

\AuthorCitation{Uehlein, M.; Weber, S. T.; Rethfeld, B.}

\address[1]{%
Department of Physics and Research Center OPTIMAS, Technische Universität Kaiserslautern,\\ 67663 Kaiserslautern, Germany
}

\corres{Correspondence: uehlein@physik.uni-kl.de} 

\abstract{
    We investigate a temperature-based model, called extended two-temperature model (eTTM), that describes the electronic non-equilibrium and its effect on energy dissipation 
    in metals after ultrashort laser excitation.
        We derive and discuss improvements in comparison to published versions of this model [E.~Carpene, Phys.~Rev.~B \textbf{2006}, 74, 024301; G.~Tsibidis, Appl.~Phys.~A \textbf{2018}, 124, 311].
        The comparison of the  
        results of the eTTM with results of the well-known two-temperature model (TTM)
        shows
        a delayed increase of the electronic temperature when being calculated with the eTTM. 
        We find a good agreement in the non-equilibrium energy distribution after absorption of photons with results from a  
        kinetic description using a Boltzmann collision term.
        The model provides a convenient tool for fast calculation of features of the non-equilibrium electrons. As an example we inspect the 
        dynamics of high-energy electrons observable in photo-electron spectroscopy and demonstrate the 
        advantage of the eTTM over the conventional two-temperature model. 
}


\begin{document}


\section{Introduction}
The high relevance of lasers in current research is undeniable. 
In particular, ultrashort pulses in the femtosecond regime are of enormous importance for materials' processing
for technical or medical applications
\cite{BaeuerleBuch11, Anisimov2002, Balling2013, Vogel2003, Vogel2005, Chichkov1996, Lu2013, Rizvi2003, Ostendorf2002, Vorobyev2013, Costache2004, Reif2010}.
The response of the solid matter on such kind of irradiation is also of great interest from the fundamental point of view~\cite{Anisimov2002, Rethfeld2004b, Rethfeld2017, Beyazit2020}.
Especially the time range of a few tens of femtoseconds after laser excitation is of importance, since intrinsic collision processes within the material take place on such ultrafast timescales. 
Due to the disturbance of the equilibrium in the electrons, no temperature is defined from the first. 
Theoretical and experimental studies are devoted to the understanding and description of the subsequent thermalization processes~\cite{Girardeau1995, Fann1992a, DelFatti2000, Baranov2014}. 
On the theoretical side, the non-equilibrium electron kinetics can be traced with various methods, e.g. 
Monte-Carlo simulations~\cite{Briones2021, Medvedev2011} 
or numerical solutions of the  
Boltzmann equation~\cite{Mueller2013PRB, Weber2019, Nenno2016}.
Many of these models are
rather complex and numerically expensive. 
Their costs and benefits may not be well-balanced, when the details of the energy distribution can be neglected.
In many cases, an approximate consideration of the laser-induced non-equilibrium  can be sufficient.
For instance, the non-equilibrium between electrons and phonons is captured in an approximative way by the  two-temperature Model (TTM), first established by Anisimov et al.~\cite{Anisimov1974}.
It is well-known, well validated on time scales of thermalized electrons \cite{Schoenlein1987, Chichkov1996}, and easily extendable~\cite{Raemer2014, Mueller2014PRB}.
Several modifications of the TTM have been proposed to capture main features of the electronic non-equilibrium~\cite{Tsibidis2018, Carpene2006, Sun1994, Maldonado2017}.
We present a slightly improved version of the model from Tsibidis~\cite{Tsibidis2018}, which was introduced as an extension of the model from Carpene~\cite{Carpene2006}. 
We investigate the description of the electronic non-equilibrium in the framework of this so-called extended two-temperature model (eTTM).\\

In this paper, we first recall the basic idea of the eTTM, describe the temporal development of the non-equilibrium subsystem, and introduce the changes made in comparison to Ref~\cite{Tsibidis2018}.
Then, we show selected results of the eTTM and compare to the corresponding results of the TTM.
We investigate the influence of the presented improvements
by showing details of the calculated non-equilibrium distribution. 
Finally, we compare the results of photon-absorption
using the TTM, eTTM and a Boltzmann calculation.
The importance of the description of the non-equilibrium becomes clear when the dynamics of the electron number density in a specific energy range above the Fermi edge is evaluated.
This quantity is accessible by time resolved two-photon photoemission spectroscopy (tr-2PPE) measurements.

\section{Theoretical Model}\label{sec:model}
\subsection{Two temperatures and a non-equilibrium system}
When a laser irradiates a metal, photons are absorbed by the electrons in the solid,
causing an increase of the electronic temperature.
Through subsequent electron-phonon collisions,
energy is transferred to the lattice, leading 
to a joint temperature of electrons and phonons.
Usually, this is described by the well-known two-temperature model (TTM)~\cite{Anisimov1974}.
It consists of two differential equations for the change of the internal energy density \(u\) of the electrons 
and the phonons, respectively, 
with the time \(t\)
\begin{linenomath*}
\begin{subequations}
\label{eq:TTM}
\begin{alignat}{9}
    &\frac{{\rm d} u\el}{{\rm d} t} &&= c\el \frac{\partial T\el}{\partial t} &&= - &&g \left(T\el-T\ph\right) &&\,+\,&&s(t)\,,\label{eq:TTM_el}\\
    &\frac{{\rm d} u\ph}{{\rm d} t} &&= c\ph \frac{\partial T\ph}{\partial t} &&= && g \left(T\el-T\ph\right)\text{,}\label{eq:TTM_ph}
\end{alignat}
\end{subequations}
\end{linenomath*}
where the index "el" stands for the electrons and "ph" for the phonons.
The change of the internal energy $u$ is expressed through the heat capacity $c$ and 
the change of the temperature~$T$. 
The laser energy enters only the electronic system and is given through 
the laser power density \(s(t)\).
To describe the above explained coupling between electrons and phonons, a term that consists of the temperature-difference of the subsystems and a proportionality factor called electron-phonon coupling parameter, \(g\)\,, appears in both equations~\cite{Kaganov1957}.
More details regarding the laser are listed in \cref{sec:laser}. 
For more information about the heat capacity, see \cref{sec:heat_cap}.\\

On short timescales, the absorption of the photons is, however, more complex. 
When the energy distribution of electrons becomes relevant, the energy density alone is not sufficient to describe the system.
In fact, only a small fraction of all electrons is able to absorb photons. 
At moderate intensities, only single photons are absorbed. 
The absorbing electrons are located in a range of one photon energy below the Fermi edge and will, after excitation, populate states in a range of one photon energy above the Fermi edge.
Due to Pauli-blocking other transitions are unlikely to occur at low temperatures. 
As a result, the Fermi distribution is strongly disturbed and no temperature can be defined for the electrons.
The excited non-equilibrium electrons thermalize back to equilibrium by collisions with other electrons and phonons.\\

\begin{figure}
    \includegraphics[width=236pt]{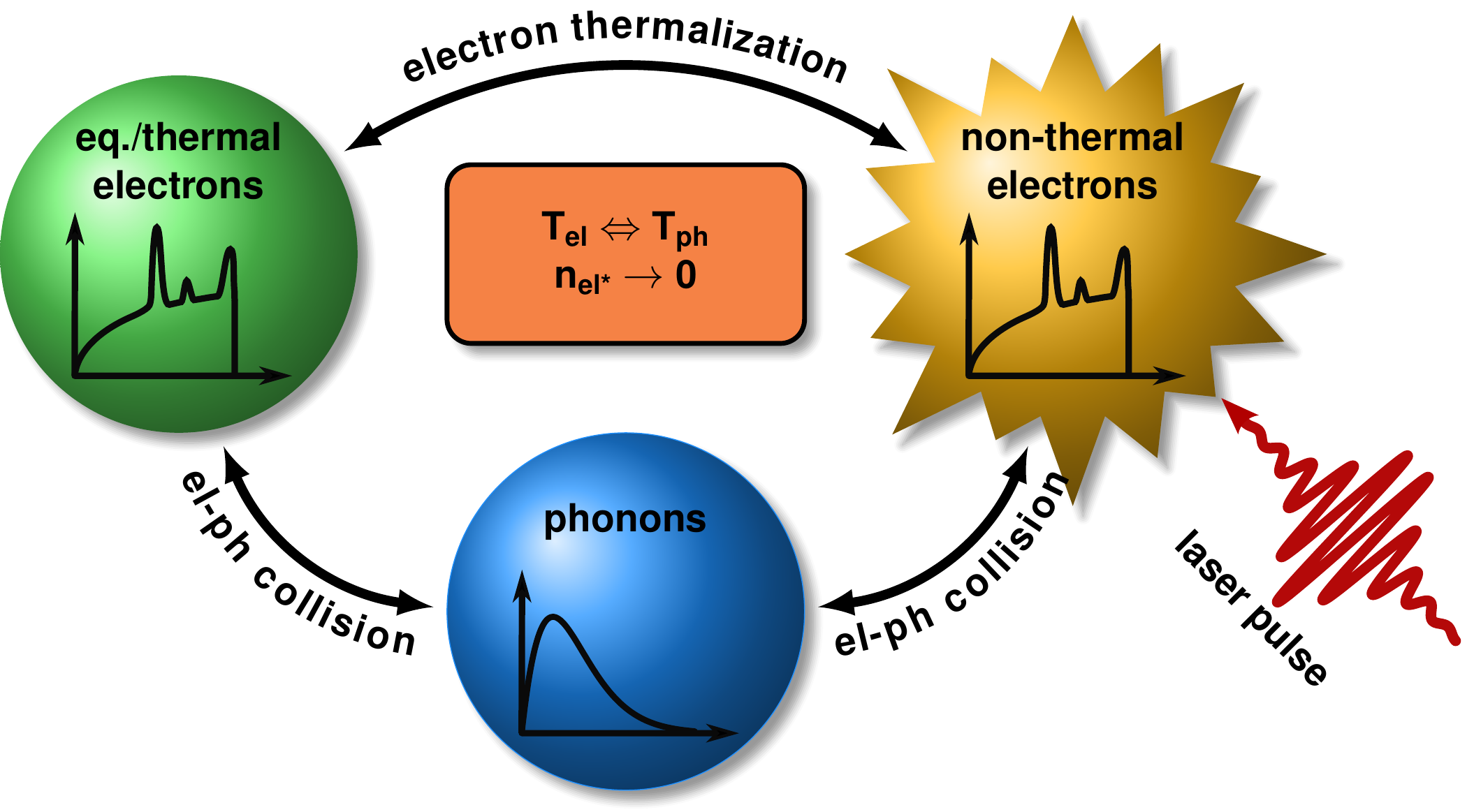}
    \caption{Schematic illustration of the subsystems and the interactions in the extended two-temperature model.
    The laser generates non-equilibrium (non-thermal) electrons. 
    They thermalize by collisions with the unaffected (thermal) electrons.
    Thus the particle density of excited electrons \(n_\text{el*}\) decreases during thermalization.
    Moreover, collisions of electrons with phonons lead to a heating of the crystal lattice.
    Both non-thermal and thermal electrons contribute to this relaxation process, which results in a joint electron and phonon temperature.
    }\label{fig:panda_carpene}
\end{figure}

To describe the laser-excited non-equilibrium and its thermalization, Carpene~\cite{Carpene2006} and later Tsibidis~\cite{Tsibidis2018} have established a model, in the following called extended two-temperature model (eTTM). 
It uses three subsystems, one phonon system like in the TTM and two electron systems.
One of these electron systems describes the high energetic, non-equilibrium electrons above the Fermi edge and the associated holes below the Fermi edge, while the other system describes the thermal background of the electrons.
The scheme of this model is shown in \cref{fig:panda_carpene}.
The laser generates non-equilibrium electrons, which are marked with a star-like shape.
The interactions between all systems are shown through arrows.
Carpene introduced this model assuming a constant density of states~(DOS) around the Fermi edge~\cite{Carpene2006}.
This gave him the opportunity to solve large parts of the model analytically.
Tsibidis improved this model by considering a more realistic DOS~\cite{Tsibidis2018}. 
As a consequence, some components of the model can only be solved numerically.
In the following, a slightly modified 
version of Tsibidis eTTM is presented.
We will discuss the improvements compared to the version of Tsibidis in \cref{sec:improvements}.

The extended two-temperature model, like the conventional TTM, consists of a differential equation for the internal energy density \(u\) of the thermal electrons and one for the internal energy density of phonons, both including the interaction with the non-thermal electrons,
\begin{linenomath*}
\begin{subequations}
\label{eq:eTTM}
\begin{alignat}{9}
    &\frac{{\rm d} u\el}{{\rm d} t} &&=- &&g \left(T\el-T\ph\right) &&\,+\,&&\frac{\partial u\elel}{\partial t}&&=:-&&\Pi_\text{el-ph}(t)&&\,+\,&&\Xi\elel(t)\text{,}\label{eq:eTTM_el}\\
    &\frac{{\rm d} u\ph}{{\rm d} t} &&= && g \left(T\el-T\ph\right) &&\,+\,&&\frac{\partial u\elph}{\partial t}&&=:&&\Pi_\text{el-ph}(t)&&\,+\,&&\Xi\elph(t)\text{,}\label{eq:eTTM_ph}
\end{alignat}
\end{subequations}
\end{linenomath*}
where the index "el" denotes the thermal and "el\(^*\)" the non-thermal electron system.
Each of the differential equations contains the electron-phonon coupling term, abbreviated here as \(\Pi_\text{el-ph}\).
The term describes the coupling between the thermalized electrons and the phonons, like in the TTM, cf.~the first terms on the right hand side of equations \mbox{(\ref{eq:TTM_el})} and (\ref{eq:TTM_ph}).
In the eTTM, the excitation term  \(s(t)\) is replaced by a coupling term of the non-thermal electrons to the thermalized electrons and phonons, called \(\Xi\elel=\frac{\partial u\elel}{\partial t}\) and \(\Xi\elph=\frac{\partial u\elph}{\partial t}\), respectively.
The \(\Xi\)-terms and the description and dynamics of the non-equilibrium system are explained in detail in the next section.

\subsection{Time evolution of the non-equilibrium system}

In order to describe the theory of this model as close as possible to its numerical implementation, time is considered as a discrete quantity.
Here \(\Delta t\) is the time step of this discretization.
When photons with energy \(E\pt\) reach the material within the time interval \([t', t'+\Delta t]\), they are absorbed and non-equilibrium electrons are generated.
These non-equilibrium electrons and the associated holes are described by the difference from the current Fermi distribution given as \(f(E) := f(t', E) := f(T\el(t'), E)\), which describes the thermalized electronic background.
It is assumed that this difference to equilibrium, the so-called excitation function, can be described by
\begin{linenomath*}
\begin{align}
    \Delta f_\text{L}(t', E) = \delta(t') \left\{f(E-E\pt) \left[1-f(E)\right]-f(E)\left[1-f(E+E\pt)\right]\right\}\,.\label{eq:stufe}
\end{align}
\end{linenomath*}
This excitation function, driven by the laser, generates non-equilibrium electrons within in the time-interval \([t', t'+\Delta t]\).
It modulates a step-like structure around the Fermi edge, which builds up when the absorption under consideration of the Pauli principle takes place~\cite{Mueller2013PRB}.
For every observed energy level \(E\), two possible processes can happen.
Electrons from the energetically lower level \(E-E\pt\) can absorb one photon, leading to an increase of the distribution at energy \(E\).
On the other hand already existing electrons at energy \(E\) can scatter into the higher level \(E+E\pt\).
The probability of the first process is proportional to 
the term \(f(E-E\pt) \left[1-f(E)\right]\) while the probability of the latter process can be expressed through \(f(E)\left[1-f(E+E\pt)\right]\).
The generated steps have a width that corresponds to the photon energy \(E\pt\).
The amplitude of the step~\(\delta\) is given by the energy conservation
\begin{linenomath*}
\begin{align}
    \int_{-\infty}^{\infty} \Delta f_\text{L}(t', E)\,D(E)\,E\,\mathrm{d}E = s(t') \Delta t \label{eq:energy}
\end{align}
\end{linenomath*}
under influence of the laser with power density \(s(t')\).
The energy-dependent \(D(E)\) denotes the density of states of the electrons.
In this work the energy zero is set to the Fermi edge, so the integration starts at negative energy.\\

At time \(t>t'\), the electrons generated in the time interval \([t', t'+\Delta t]\) thermalize by collisions with electrons and phonons.
According to Ref.~\cite{Carpene2006, Tsibidis2018}, the dynamics of the excitation function can be described by
\begin{linenomath*}
\begin{align}
    \Delta f'(t, t', E) = \Delta f_\text{L}(t', E) \exp\left(-\frac{t-t'}{\tau\elel}-\frac{t-t'}{\tau\elph}\right)\,,\label{eq:delta_f}
\end{align}
\end{linenomath*}
using energy relaxation times \(\tau\elel,~\tau\elph\).
We use an energy- and temperature-dependent electron-electron relaxation time
\begin{linenomath*}
\begin{align}
    \tau\elel(E, T) = \tau_0 \frac{E_\text{F}^2}{(E-E_\text{F})^2+(\pi k_\text{B} T_\text{el})^2}\,,\label{eq:relax_elel}
\end{align}
\end{linenomath*}
with pre-factor \(\tau_0\)\,.
It is a simplified version from Fermi-liquid theory~\cite{Mueller2013PRB}.
In contrast to the 
energy dependent relaxation time used in Ref.~\cite{Tsibidis2018}, equation~\mbox{(\ref{eq:relax_elel})} prevents a singularity at the Fermi edge.
The electron-phonon relaxation time \(\tau\elph\) is calculated according to Ref.~\cite{Carpene2006, Groeneveld1995}.\\
More details about the parameters applied are listed in \cref{sec:relax_param}.\\

For various investigations, it is useful to calculate the present distribution that describes all electrons, thermalized and non-thermalized.
The temporal dynamics of the particle number in a specific energy range is also an interesting quantity, because it is comparable to experimental photoemission results (see~\cref{sec:experiment}).
The present distribution
\begin{linenomath*}
\begin{align}
    f^\text{tot}(t, E) = f(t, E) + \Delta f(t, E) = f(t, E) + \sum_{\substack{t' = 0,\\  t' += \Delta t}}^{t} \Delta f'(t, t', E)\text{.} \label{eq:f_tot}
\end{align}
\end{linenomath*}
is given by adding the present corresponding Fermi distribution \(f(t, E)\) and the sum of the thermalizing excitation functions (\cref{eq:delta_f}) for all previous time intervals.\\

From \cref{eq:delta_f}, the total energy transfer from the system of the non-equilibrium electrons to the equilibrium systems can be calculated.
At time \(t\), the non-equilibrium electrons generated in the time interval \([t',t'+\Delta t]\) provide the contribution
\begin{linenomath*}
\begin{align}
    \frac{\partial u'(t, t')}{\partial t}    &= -\frac{\partial}{\partial t}\int_{-\infty}^{\infty} \Delta f'(t, t', E)\,D(E)\,E\,\mathrm{d}E \label{eq:u_ettm}\\
    &= - \int_{-\infty}^{\infty} \left(-\frac{1}{\tau\elel}-\frac{1}{\tau\elph}\right) \Delta f_\text{L}(t', E) \exp\left(-\frac{t-t'}{\tau\elel}-\frac{t-t'}{\tau\elph}\right)\,D(E)\,E\,\mathrm{d}E \nonumber\\
    &=: \frac{\partial u'\elel(t, t')}{\partial t} + \frac{\partial u'\elph(t, t')}{\partial t} =: \Xi\elel'(t, t') + \Xi\elph'(t, t')\,. \nonumber
\end{align}
\end{linenomath*}
The total energy rate \(\frac{\partial u'(t, t')}{\partial t}\) that enters the equilibrium systems can be split up into one contribution entering the equilibrium electrons \( \Xi\elel'(t, t')\) and one contribution entering the phonons \(\Xi\elph'(t, t')\).\\
To consider the non-equilibrium electrons generated in all previous time steps, it is necessary to sum over them. 
This results in the energy rates
\begin{linenomath*}
\begin{align}
    \Xi_\text{el$^*$-el/ph}(t) := \frac{\partial u_\text{el$^*$-el/ph}(t)}{\partial t} = \sum_{\substack{t' = 0,\\  t' += \Delta t}}^{t} \Xi_\text{el$^*$-el/ph}'(t, t') =  \sum_{\substack{t' = 0,\\ t' += \Delta t}}^{t} \frac{\partial u_\text{el$^*$-el/ph}'(t, t')}{\partial t} \label{eq:gesamt_u}
\end{align}
\end{linenomath*}
that enter the differential equation system (\ref{eq:eTTM}).
The summation is done in steps of \(\Delta t\) to match the time intervals.

\subsection{Improvements as compared to the work of Tsibidis~\cite{Tsibidis2018}}\label{sec:improvements}
The model presented in the previous section is a slightly improved version of the eTTM published in Ref.~\cite{Tsibidis2018}.
Besides the enhanced electron-electron relaxation time (see \mbox{\cref{eq:relax_elel}}), also the integration limits are improved.
In contrast to the eTTMs published so far~\cite{Carpene2006, Tsibidis2018}, the entire energy range is considered here for the integrations in equations (\ref{eq:energy}) and (\ref{eq:u_ettm}).
The previous models only integrated over the range \([E_\text{F}-E\pt,~E_\text{F}+E\pt]\) around the Fermi edge \(E_\text{F}\), which entails an nonphysical limitation of absorption at temperatures above absolute zero.\\

\section{Results}
The model is applied to aluminum as an example material. If not otherwise specified, the parameters for the material and for the Gaussian laser pulse are listed in \cref{sec:anhang}. In the following, all time-dependent figures show the temporal shape of the laser pulse in grey.\\

The description given here is valid for a thin metal film, because any transport through the material was neglected and homogeneous heating was assumed.
Tsibidis considered fundamentally the transport of the equilibrium electrons by neglecting any transport effects of the non-equilibrium electrons~\cite{Tsibidis2018}.
However, their influence should be very relevant~\cite{Rethfeld2017, Knoesel1998, Beyazit2020}.
To the best of our knowledge, there is no known description considering non-equilibrium energy transport in temperature-based models like the eTTM. 

\subsection{Temperature and energies}
    In the following, we investigate the temperatures and energies of the subsystems, as well as the energy transfer between them.
    Comparing results of the eTTM to results of the TTM, we highlight
    the influences of non-equilibrium electrons.\\
    
    \begin{figure}
        \includegraphics{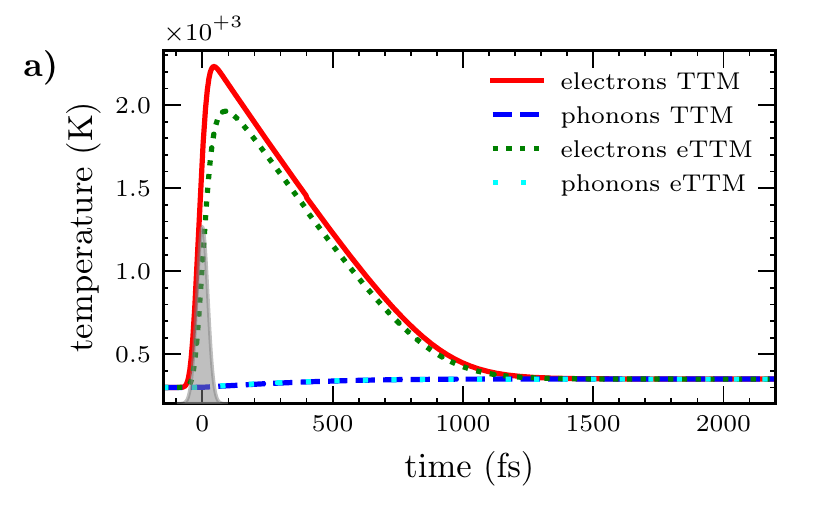}
        \caption{Temperatures of electrons and phonons. 
            The results of a TTM simulation are compared to the results of an eTTM simulation.
            The shape of the laser pulse is sketched in gray.
            It is visible that the maximum temperature of the electrons in the TTM simulation is higher than the maximum electron temperature obtained with the eTTM simulation.
            At this stage, part of the energy is kept in the non-thermal electrons of the eTTM.}\label{fig:temp_ttm}
    \end{figure}
    
    \Cref{fig:temp_ttm} shows the dynamics of the temperatures of electrons and phonons obtained with both models.
    Please notice, that no temperature can be assigned to the non-equilibrium electrons.
    Therefore, only the temperature of the equilibrium background is shown for the eTTM.
    The electron temperatures of both models show the initial rise through the irradiating laser pulse, as well as, the relaxation to a joint temperature of electrons and phonons.
    As it is expected and desired, there is no major difference between the temperatures of a TTM to an eTTM simulation.
    Including the non-equilibrium electrons in the eTTM leads to a lower maximum electron temperature in comparison to the TTM temperature. 
    Moreover, the electrons are heated more slowly and a bit longer.
    This is caused by the energy of the laser is not being directly supplied to the thermal electrons.
    Due to the indirect heating through the non-equilibrium electrons, the increase in temperature is distributed over a larger time range.
    This will be discussed in the next paragraph by observing the energy dynamics of the subsystems.
    The relaxation process of both models is the same.
    The phonon temperature with the eTTM differs only slightly from the TTM phonon temperature.\\

    \begin{figure}
        \includegraphics{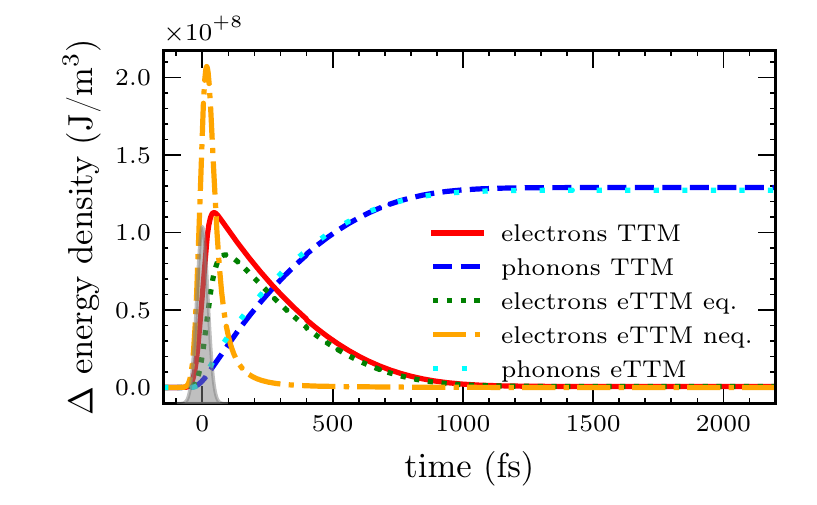}
        \caption{Energy density of all subsystems of TTM and eTTM in dependence on time,
        plotted as difference to the initial values.
            Due to the connection between energy density and temperature, the dynamics follow the temperature dynamics (\cref{fig:temp_ttm}).
            Additionally the energy density of the non-equilibrium electrons in the eTTM is shown in the yellow dashed-dotted line.
            It is clearly visible that non-equilibrium electrons are generated directly by the laser and thermalize in the next hundreds of femtoseconds.
        }\label{fig:energies_TTM}
    \end{figure}
    In contrast to the temperature, the energy density of the non-equilibrium electrons is well defined.
    This allows us to compare the dynamics of the non-equilibrium system of the eTTM with that of the equilibrium systems of both models.
    \Cref{fig:energies_TTM} depicts the energy densities of all subsystems of the models in dependence on time. 
    They are plotted as a difference to their initial value before the laser pulse, to compare the dynamics of the different quantities.
    The change of the equilibrium electron and phonon temperatures are connected to the energy gain via the heat capacity.
    Therefore the energy dynamics of those systems is qualitative equal to the behavior explained before (\cref{fig:temp_ttm}).
    Due to the higher phonon heat capacity, the phonons have gained much more energy than the electrons at the final stage. 
    The influence of the non-equilibrium, that could not be shown in \cref{fig:temp_ttm}, is clearly visible, here.
    The rise of the energy of the non-equilibrium electrons with the laser pulse can be observed.
    Within \SI{200}{\fs} after the laser pulse the thermalization is completed.
    This corresponds to thermalization times extracted from two-photon photoemission experiments~\cite{Beyazit2020}. 
    The peak energy contained in the non-equilibrium electrons is more than two times larger than the peak energy in the equilibrium electron system.
    This is due to the fact that the laser delivers energy to the non-equilibrium electrons faster than it is transferred to the equilibrium electrons by thermalization.\\
    \begin{figure}
        \includegraphics{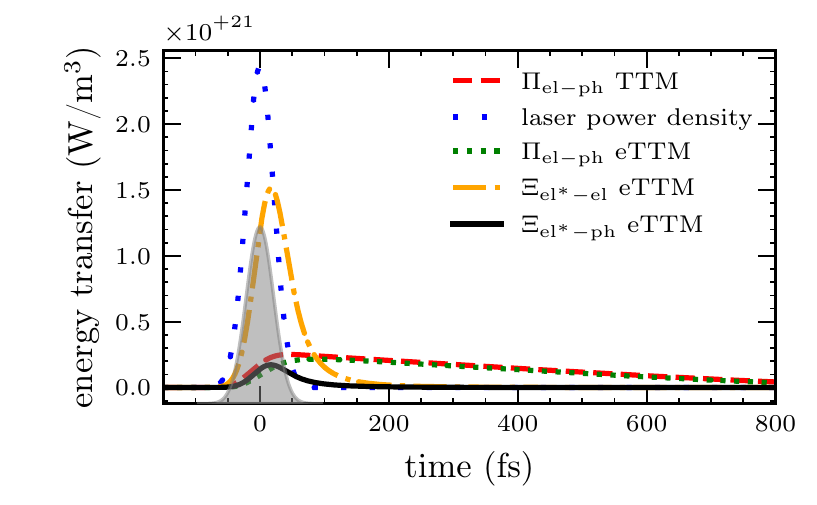}
        \caption{
            Dynamics of the energy density change in dependence of time.
            The notation corresponds to the definitions introduced in equations (\ref{eq:eTTM}) and (\ref{eq:TTM_el}).
            To compare TTM and eTTM, the laser power density \(s(t)\) and the electron-phonon coupling term of a TTM simulation are shown.
            The equilibrium coupling terms (\(\Pi_\text{el-ph}\)) of TTM and eTTM are very similar.
            The energy transferred by the laser power density in the TTM is split up to the \(\Xi\)-terms in the eTTM.
            These terms heat the equilibrium electrons considerably longer than the laser power density in the TTM.
            The coupling of the non-equilibrium electrons to the phonons (\(\Xi\elph\)) is much smaller than the coupling of the non-equilibrium electrons to the equilibrium electrons (\(\Xi\elel\)).
        }\label{fig:dudt_TTM}
    \end{figure}
    
    To investigate the coupling strengths between the subsystems, \cref{fig:dudt_TTM} shows the energy transfer between all of them in dependence of time. 
    The curves correspond to the terms from equations (\ref{eq:eTTM}) together with the laser source term and electron-phonon coupling term in (\ref{eq:TTM_el}).
    The \(\Pi_\text{el-ph}\)-terms describe the coupling between the equilibrium electrons and the phonons, while the \(\Xi\)-terms express the coupling between the non-equilibrium electrons and the equilibrium electrons or phonons, respectively.
    In general, the coupling between electronic and phononic subsystems is weaker than the coupling between the electrons.
    In the case of the TTM, the energy from the laser that is transferred to the equilibrium electrons is expressed through the laser power density \(s(t)\).
    Regarding the eTTM, the electrons are not directly heated by the laser pulse.
    Thus, the laser term is replaced  by the \(\Xi\)-terms.
    However, the integral over time of both \(\Xi\)-couplings summed up, has to equal the integral of the laser power density.
    Apart from small deviations discussed in \cref{sec:res_improvements}, this energy conservation is fulfilled.
    While the laser power density heats the electrons exactly during the laser pulse, the \(\Xi\)-terms heat the equilibrium electrons over a longer time range.
    The coupling of the equilibrium electrons to the phonons (\(\Pi_\text{el-ph}\)) is nearly identical in the TTM and the eTTM.\\

\end{paracol}
    \begin{figure}
        \widefigure
        \includegraphics{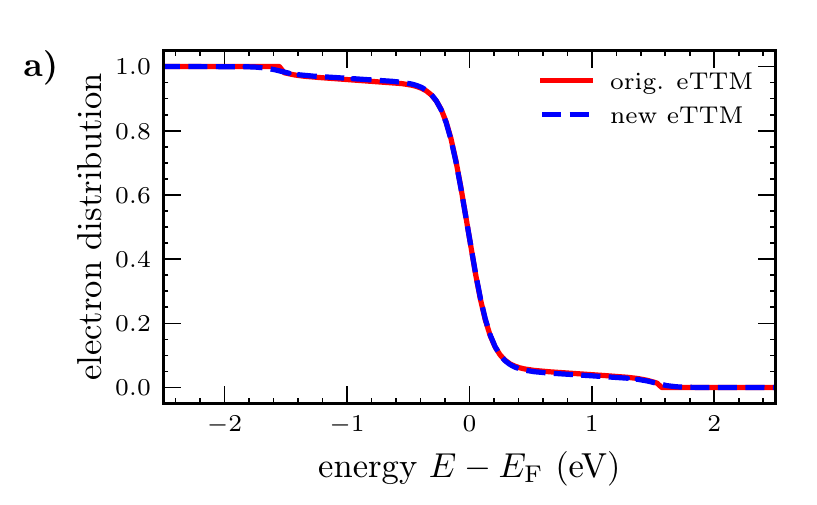}
        \hspace{.5cm}
        \includegraphics{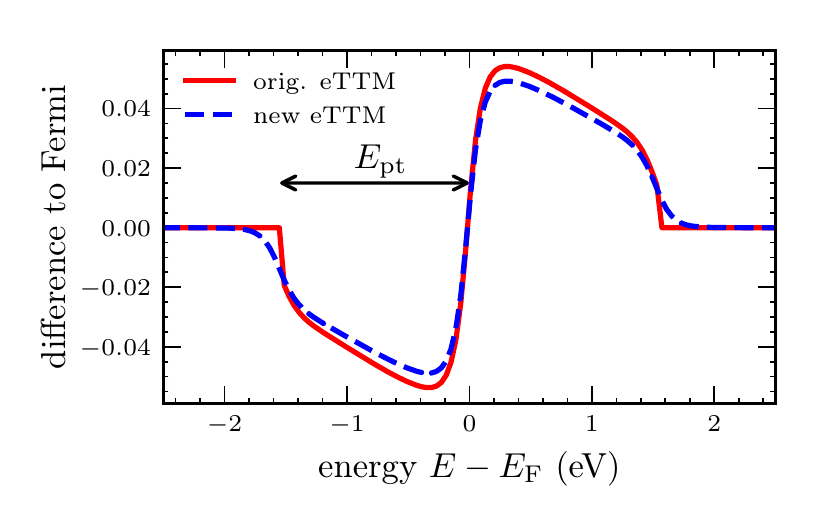}
        \caption{Electron distributions at the maximum of the laser pulse.
            The original eTTM by Tsibidis~\cite{Tsibidis2018} is compared to the new and modified version presented in this paper. 
            Figure a) shows the electron distribution, figure b) the difference to the equilibrium background. 
            The step is visible and has a width that equals the photon energy \(E\pt = \SI{1.55}{\electronvolt}\), like it was explained in the theory part of the model (see~\cref{sec:model}).
            While the differences in total are small (see a)), the new model prevents the unphysical kink at the edges of the laser-disturbed energy range by distributing the energy of the laser over a wider energy range (see b)).
        }\label{fig:f_al}
    \end{figure}
    \begin{paracol}{2}
    \ifthenelse{\equal{\@status}{submit}}{\linenumbers}{}
    \switchcolumn

\subsection{Influence of the improvements regarding the integration range} \label{sec:res_improvements}

    \Cref{fig:f_al} shows the electron distribution at the maximum of the laser pulse.
    The results of a simulation with the original model~\cite{Tsibidis2018} are compared to the model presented in this work (compare~\cref{sec:improvements}).
    In figure a), the whole non-equilibrium electron distribution is shown~(\cref{eq:f_tot}), while figure b) displays the difference to the electronic background, which is described by a Fermi distribution.
    This difference is described by \(\Delta f(t, E)\) from \cref{eq:f_tot} and characterizes the system of non-equilibrium electrons.
    The curves are almost the same.
    Figure b) points out the differences in more detail.
    In both curves, the step that describes the absorption of photons is observable.
    The width of the step corresponds to the photon energy \(E\pt = \SI{1.55}{\electronvolt}\), as it is expected from the mechanism explained in \mbox{\cref{sec:model}}.
    Electrons in occupied states below the Fermi edge absorb photons and are excited to unoccupied states above the Fermi edge.
    This results in a non-equilibrium distribution, which shows approximately the curve known from kinetic simulations (compare \ref{sec:boltzmann}).
    
    The original eTTM (red, solid line) has a clearly visible kink at the outer edges of the step.
    This is an artifact of the numeric limitations of the integrated energy range, one photon energy below and above the Fermi edge.
    If we allow to distribute the energy over the whole energetic range, as it is done for the simulation displayed with the blue dashed line, the kink is prevented.
    For the same stored energy, the height of the step is lower in the blue dashed curve.
    For energies close to the Fermi edge the height in \cref{fig:f_al} b) is larger than for higher energies.
    This is due to the non-zero temperature of the Fermi distribution.
    At \SI{0}{\kelvin}, the step of excitation (see \cref{eq:stufe}) would be straight.
    Contrary than expected from results of kinetic models, the shape of the DOS does not play a role here.
    As can be seen in \cref{eq:stufe}, the DOS enters only the amplitude \(\delta(t')\) but not the shape of the step.
    This is a simplification of the eTTM, which will be discussed in more detail in \cref{sec:boltzmann}.\\
    
    \end{paracol}
    \begin{figure}
        \widefigure
        \includegraphics{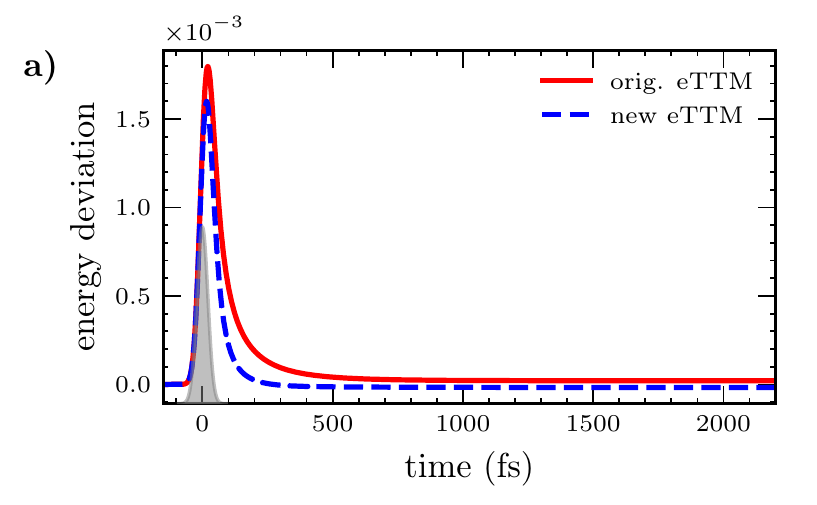}
        \hspace{.5cm}
        \includegraphics{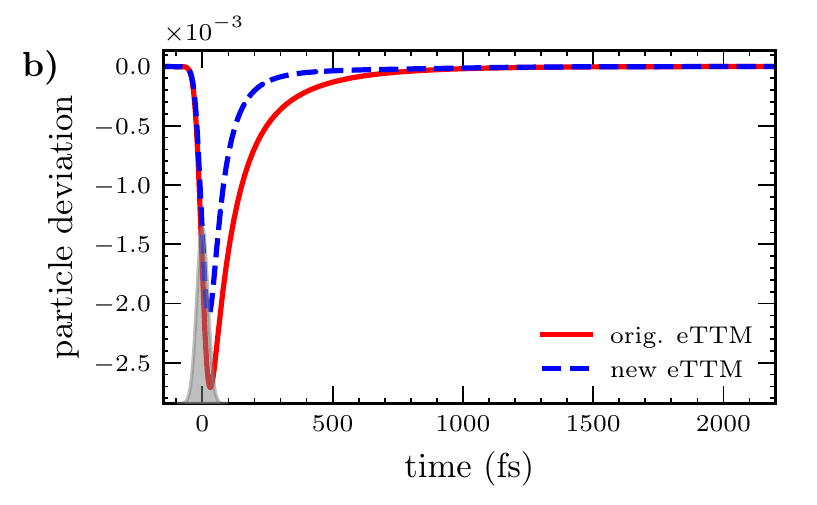}
        \caption{Deviation from conservation of a) energy and b) particles, according to \cref{eq:dev}. 
            The results of the original eTTM by Tsibidis~\cite{Tsibidis2018} is compared to the new and modified version presented in this paper. 
            We achieved a weaker deviation of conservation of energy and particles.
        }\label{fig:cons_al}
    \end{figure}
    \begin{paracol}{2}
    \ifthenelse{\equal{\@status}{submit}}{\linenumbers}{}
    \switchcolumn

    The motivation for the improvement of the integrated energy range becomes clear in \cref{fig:cons_al}.
    It shows the deviation from the conservation of the energy density (fig. a)) and particle density (fig. b)) of two eTTM simulations over time. 
    The original eTTM presented by Tsibidis in Ref.~\cite{Tsibidis2018} is compared to the model presented here. 
    The deviation
    \begin{linenomath*}
    \begin{align}
        X_\text{dev}^\text{norm}(t) = \frac{x(t)-x(t=0)}{x(t=0)} \label{eq:dev}
    \end{align}
    \end{linenomath*}
    is calculated through the difference from the start value, normalized to the start value.
    The variable \(x\) represents either the total energy or particle density in the system.
    The total energy density is calculated by adding up the energy densities of the three subsystems and subtracting the energy input of the laser at each time step.
    Both models exhibit the same qualitative behavior of the deviations.
    The energy and particles are not completely conserved.
    This is in contrast to the TTM which would yield a perfect zero-line in \mbox{\cref{fig:cons_al}} (not shown).
    During the laser pulse an increase of the energy in the system and a decrease of the particle density can be observed.
    With the thermalization of the non-equilibrium electrons, the energy deviation (\cref{fig:cons_al} a)) decreases to a level close to zero. 
    Generally, only deviations in the per mille range occur here.
    Also the particle conservation shows deviations in the per mille range, which are highest during the laser pulse and recover with during thermalization, see \cref{fig:cons_al} b).
    In both cases, the above-introduced improvements lead to a weaker violation of the conservation law and a faster recovery to fulfilled conservation. 
    This improvement is a consequence of the expansion of the energy integration range to the complete energy spectrum.
    The remaining deviation is probably due to the neglection of the DOS in the eTTM, which influences absorption, as kinetic simulations with Boltzmann collision integrals have shown~\cite{Mueller2013PRB, Weber2019}. 
    We compare results of both descriptions in 
    the next section.
    
\subsection{Comparison to kinetic description} \label{sec:boltzmann}
    As stated above, simulations based on the Boltzmann equation~\cite{Rethfeld2002, Mueller2013PRB, Weber2019, Nenno2016} describe the laser-induced non-equilibrium more accurately, allowing therefore to study detailed features of the excited electron distribution.
    Such kinetic simulations are, however, numerically rather expensive.
    The eTTM represents a compromise between purely temperature-based models and a full kinetic description.
    
    For comparison, we have calculated the excited electron distribution with the electron-ion-photon collision term of a Boltzmann simulation according to Ref.~\cite{Mueller2013PRB}, and with the eTTM and TTM, respectively.
    In all three calculations, the excitation has been performed with a laser pulse of the same wavelength. 
    The material has been initially at room temperature (\SI{300}{\kelvin}). After excitation, the electrons in the TTM have a temperature of \SI{36418}{\kelvin}. 
    The same energy density has been introduced into the electrons in the Boltzmann kinetic simulation as well as into the non-equilibrium electrons of the eTTM. 
    In both cases this was done during the first timestep, it means for the case of the eTTM that \(s(t\!=\!0)\neq0\text{ and }s(t\!>\!0)=0\) in \cref{eq:energy}. 
    This can be interpreted as an excitation with a $\delta$-like pulse shape, which isolates the influence of the absorption from thermalization effects.
    In this case, the eTTM equations are simplified in a way that the sums in \cref{eq:f_tot} and  \mbox{\cref{eq:gesamt_u}} vanish.
    
    \end{paracol}
      \begin{figure}
        \widefigure
        \includegraphics{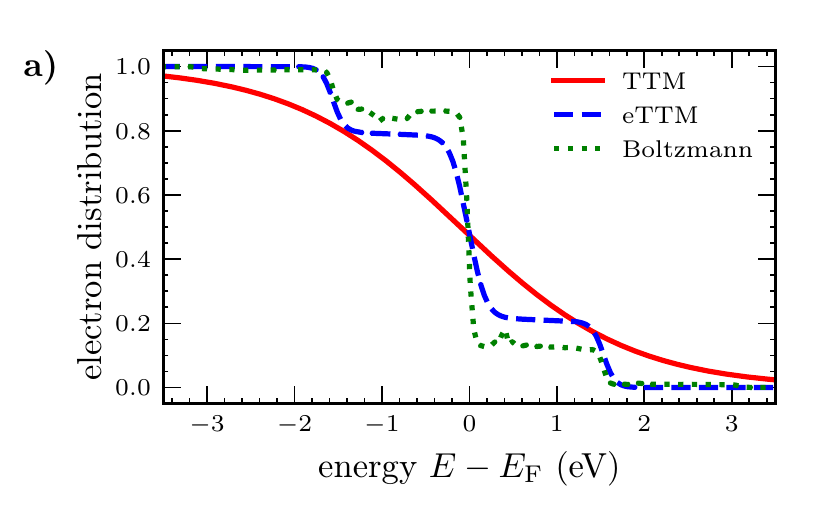}
        \hspace{.5cm}
        \includegraphics{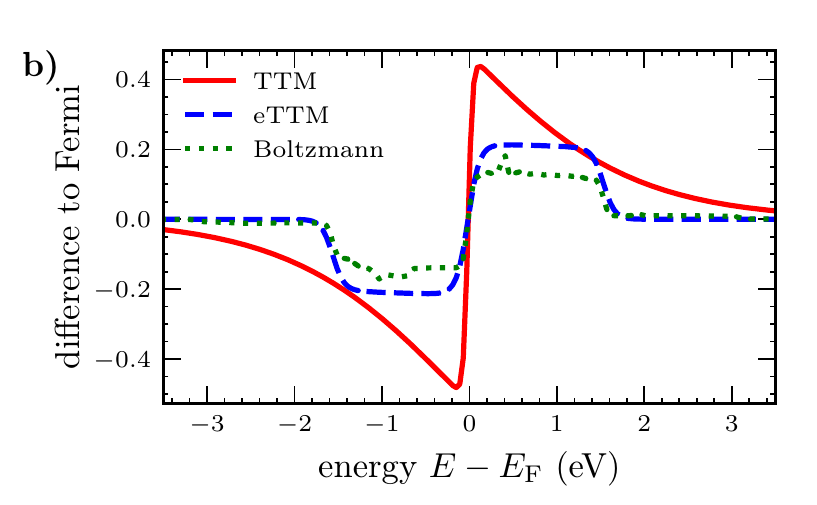}
        \caption{Electron distribution directly after a pulse with a \(\delta\)-peaked shape.
            Figure a) shows the distributions with a TTM simulation, an eTTM simulation and a simulation using the Boltzmann collision term for the electron-ion-photon interaction for the same absorbed energy density.
            The TTM results in a Fermi distribution with elevated temperature.
            In contrast, the distribution obtained from the eTTM shows two steps indicating the non-equilibrium.
            The eTTM and the Boltzmann simulation result in similar distributions.
            In figure b) the difference to the initial distribution is shown.
            The difference between the TTM and the eTTM is striking. 
            }\label{fig:rect}
    \end{figure}
    \begin{paracol}{2}
    \ifthenelse{\equal{\@status}{submit}}{\linenumbers}{}
    \switchcolumn

    In \cref{fig:rect}a) the distributions directly after the irradiation are depicted.
    \mbox{\Cref{fig:rect}b)} shows the difference of these distributions to the initial Fermi distribution at room temperature. 
    For the eTTM, this difference
    equals the excitation function \cref{eq:stufe}.
    Both, eTTM and Boltzmann simulation, show a  similar step-like structure.
    In contrast,  the TTM approach shows no steps in \cref{fig:rect}, since here a Fermi distribution of elevated temperature represents the excited distribution. 
    The main step of the Boltzmann-simulated result is less pronounced than that of the eTTM. 
    However, the Boltzmann simulation reveals a second small step,
    which is due to two-photon absorption included in this simulation. 
    Furthermore, the  Boltzmann-simulated result shows clear features of the DOS imprinted on the distribution.
    In contrast, the eTTM does not include two-photon absorption and is not able to describe the influence of the DOS on the excitation. 
    Please note that the steps of eTTM and Boltzmann simulation in \mbox{\cref{fig:rect}} are more rectangular, i.e. less sloping, than those in \mbox{\cref{fig:f_al}}.
    The initial electron temperature and thus the temperature of the thermalized electrons in the eTTM for the calculations of \cref{fig:rect} is \SI{300}{\kelvin}, while for simulations underlying \cref{fig:f_al} the temperature of the thermal electrons at the maximum of the laser pulse 
    is around \SI{1500}{\kelvin}.
    This emphasizes the above-explained dependence of the excitation function from the thermal electron temperature and also shows an effect of beginning thermalization during irradiation.
    
    Altogether, we conclude that the absorption behavior can be simulated with the help of the eTTM in a simplified form but in good agreement with the Boltzmann collision term.

\subsection{Connection to experiments}~\label{sec:experiment}
    \begin{figure}
        \includegraphics{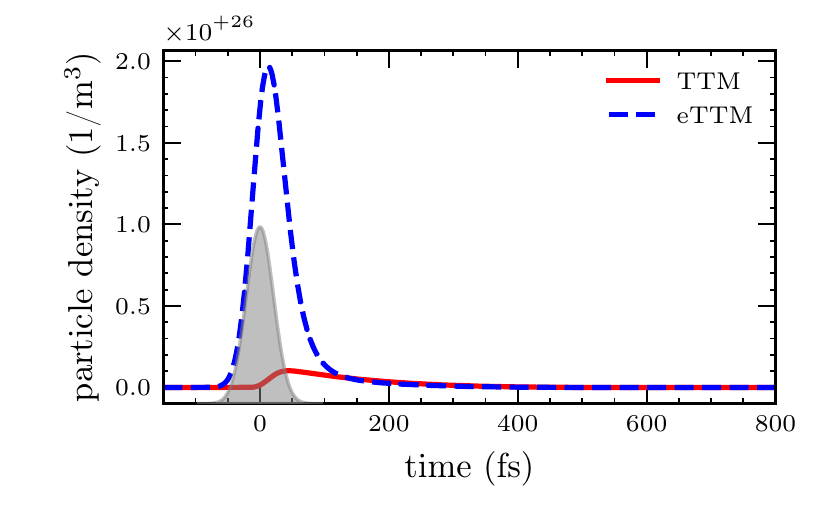}
        \caption{Dynamics of the particle density integrated over the energy range \SI{1}{\electronvolt} to \SI{1.5}{\electronvolt} above the Fermi edge. The results of a TTM simulation is compared to a eTTM simulation. 
        In the eTTM, a significant amount of high energetic electrons is generated directly with the laser pulse and thermalizes about \SI{50}{\fs} delayed to the laser.
        In contrast, the particle density in the TTM only rises delayed and much weaker.
        }\label{fig:particle_dyn}
    \end{figure}
    For many experiments the energy-resolved occupation is of interest.
    For instance, photo-electron spectroscopy (PES) is a widely used tool to investigate ultrafast processes in microscopic solid state physic.
    With the time resolved two-photon PES (tr-2PPE) it is possible to observe the dynamics of high energy electrons~\cite{Beyazit2020, Bauer2015}.
    An additional energy filter is used for this purpose.
    The eTTM can deliver important information for such kinds of experiments via the particle density in a specific energy range in dependence of time.\\
    
    In \cref{fig:particle_dyn}, the particle number density of electrons in the interval between \SI{1}{\electronvolt} and \SI{1.5}{\electronvolt} above the Fermi edge is shown. 
    A TTM simulation is compared with the eTTM.
    The results of the two models show a drastic difference.
    Directly at the beginning of the laser pulse, the number of particles in the given energy window rises sharply when calculated with the eTTM,
    and falls again rapidly after the end of the laser pulse.
    The same behavior could be observed in experiments~\cite{Beyazit2020}.
    In the TTM, the electron density and thus the prospected yield of a photoemission experiment increases mainly during the maximum of the laser pulse and decreases more slowly. 
    The peak of the partial number density in the TTM is about ten times smaller than in the eTTM.

\section{Summary}
We have recalled the extended two-temperature model as introduced by Carpene~\cite{Carpene2006} and later Tsibidis~\cite{Tsibidis2018}.
We have shown the establishment and thermalization of the electronic non-equilibrium 
in comparison to the TTM.
Our extension of the energy range in the integration in comparison to Tsibidis' version of this model improves the energy- and particle conversation.
The eTTM reproduces the typical step-like shape in the distribution function for the laser-excited non-equilibrium distribution.
As we have seen from the similarities to the full kinetic description in the absorption behavior,  the eTTM has the potential to substitute full Boltzmann calculations.
This may be particularly useful in order to obtain an impression of the influence of non-equilibrium effects in dedicated ultrafast experiments, as have been suggested for the example of photoemission spectroscopy. 
Further comparisons, e.g.~with the temporal evolution of the results obtained with full Boltzmann collision terms, are ongoing. 
Already now we can conclude that 
the extended two-temperature model is a useful tool to describe the electronic non-equilibrium in a numerically favorable way.

\vspace{6pt} 



\authorcontributions{Conceptualization, M.U., B.R. and S.T.W.; methodology, M.U.; software, M.U. and S.T.W.; validation, S.T.W. and B.R.; resources, B.R.; writing---original draft preparation, M.U.; writing---review and editing, B.R. and S.T.W.; visualization, M.U.; supervision, B.R.; project administration, B.R.; All authors have read and agreed to the published version of the manuscript.}

\acknowledgments{Many thanks to Christopher Seibel for his support in programming the simulation.}

\conflictsofinterest{The authors declare no conflict of interest.}



\appendixtitles{yes} 
\appendixstart
\appendix
\section{Simulation and material parameters}\label{sec:anhang}
In this section the setup of the laser and the parameters for aluminum are listed.
\subsection{laser setup}\label{sec:laser}
A laser pulse with Gaussian profile is used. 
The intensity 
\begin{linenomath*}
\begin{align}
    I(t) = \sqrt{\frac{4\ln(2)}{\pi}} \frac{\Phi}{\tau_\text{p}}\exp\left[-4\ln(2)\left(\frac{t}{\tau_\text{p}}\right)\right]
\end{align}
\end{linenomath*}
at time \(t\) can be calculated with the fluence \(\Phi=\SI{15}{\joule\per\square\meter}\) and the full width at half maximum (FWHM), \(\tau_\text{p}=\SI{50}{\fs}\).
With this the laser power density \(s(t)= (1-R) \alpha I(t)\) is given. Here \mbox{\(\alpha = \SI{1.3123e8}{\per\meter}\)} is the absorption coefficient and \(R=\SI{0.8682}{}\) the reflectivity of aluminum~\cite{Rakic1995}. We assume a monochromatic pulse with photon energy \(E\pt = \SI{1.55}{\electronvolt}\).

\subsection{heat capacities}\label{sec:heat_cap}
The connection between electron distribution \(f(T_\text{el}, E)\) and the particle density
\begin{linenomath*}
\begin{align}
    n_\text{el} = \int_{-\infty}^\infty f(T_\text{el}, E)\,D(E)\,{\rm d}E
\end{align}
\end{linenomath*}
respectively the internal energy density
\begin{linenomath*}
\begin{align}
    u_\text{el} = \int_{-\infty}^\infty f(T_\text{el}, E)\,D(E)\,E\,{\rm d}E
\end{align}
\end{linenomath*}
is well-known from solid state physics text books.
From this, the electron heat capacity
\begin{linenomath*}
\begin{align}
    c\el = \dv{u}{T} = \int_{-\infty}^\infty \dv{f(T_\text{el}, E)}{T}\,D(E)\,E\,{\rm d}E
\end{align}
\end{linenomath*}
is calculated directly from the DOS \(D\) of the electrons~\cite{ibach, hunklinger}. 
The DOS was calculated with density functional theory and is taken of Ref.~\cite{Lin2008}.
The energy zero is set to the Fermi edge, so the integration starts at negative energies.\\
The phonon heat capacity 
\begin{linenomath*}
\begin{align}
    c\ph = 3 n\ph k_\text{B} = 3 \frac{\rho_\text{V}N_\text{A}}{M} k_\text{B} = \SI{2.5e6}{\joule\per\kelvin\per\cubic\meter}
\end{align}
\end{linenomath*}
is assumed to be constant according to the Dulong-Petit law. Here the particle density \(n\ph\) is calculated with the density \(\rho_V = \SI{2.7e6}{\gram\per\cubic\meter}\) and the molar mass \mbox{\(M = \SI{26.98}{\gram\per\mol}\)}~\cite{CRC2005}.

\subsection{relaxation time parameters}\label{sec:relax_param}
Here, the relaxation time parameters used in equations (\ref{eq:delta_f}) and (\ref{eq:relax_elel}) are listed.
From Fermi liquid theory, the pre-factor
\begin{linenomath*}
\begin{align}
    \tau_0 = \frac{128}{\sqrt{3}\pi^2\omega_\text{P}} = \SI{0.329}{\fs}
\end{align}
\end{linenomath*}
is calculated with the plasma frequency \(\omega_\text{P} = \frac{\SI{14.98}{\electronvolt}}{\hbar}\)~\cite{Carpene2006, Pines, Rakic1998}.
Note that this pre-factor of \cref{eq:relax_elel} results in thermalization times up to picosecond timescale at the Fermi edge.\\
The electron-phonon relaxation time calculates according to Ref.~\cite{Carpene2006, Groeneveld1995} with the data from Ref.~\cite{Ashcroft,Kittel} to \(\tau\elph = \SI{299.22}{\fs}\).

\subsection{miscellaneous}
We assume the electron-phonon coupling parameter to be constant and use a value of \(g = \SI{2.6148e17}{\joule\per\kelvin\per\second\per\cubic\meter}\) taken from Ref.~\cite{Lin2008} at \SI{1000}{\kelvin}.\\
A FCC crystal with lattice constant \SI{4.05e-10}{\meter} is assumed~\cite{Lin2008}.
\end{paracol}
\reftitle{References}


\externalbibliography{yes}
\bibliography{bibfile/all.bib}

\end{document}